\documentclass[9pt]{acm_proc_article-sp}
\usepackage[hyphens]{url}
\usepackage{graphicx}
\usepackage{booktabs} 

\usepackage[T1]{fontenc}
\usepackage{lmodern}
\usepackage{amssymb,amsmath}
\usepackage{ifxetex,ifluatex}
\usepackage{fixltx2e} 
\IfFileExists{upquote.sty}{\usepackage{upquote}}{}
\ifnum 0\ifxetex 1\fi\ifluatex 1\fi=0 
  \usepackage[utf8]{inputenc}

\else 
  \usepackage{fontspec}
  \ifxetex
    \usepackage{xltxtra,xunicode}
  \fi
  \defaultfontfeatures{Mapping=tex-text,Scale=MatchLowercase}
  
\fi

\IfFileExists{microtype.sty}{\usepackage{microtype}}{}
\usepackage{color}
\usepackage{fancyvrb}

\DefineVerbatimEnvironment{Highlighting}{Verbatim}{commandchars=\\\{\}}
\usepackage{framed}
\definecolor{shadecolor}{RGB}{248,248,248}
\newenvironment{Shaded}{\begin{snugshade}}{\end{snugshade}}
\newcommand{\KeywordTok}[1]{\textcolor[rgb]{0.13,0.29,0.53}{\textbf{{#1}}}}

\newcommand{\OtherTok}[1]{\textcolor[rgb]{0.56,0.35,0.01}{{#1}}}

\newcommand{\NormalTok}[1]{{#1}}
\ifxetex
  \usepackage[setpagesize=false, 
              unicode=false, 
              xetex]{hyperref}
\else
  \usepackage[unicode=true]{hyperref}
\fi
\hypersetup{breaklinks=true,
            bookmarks=true,
            pdfauthor={},
            pdftitle={An introduction to Docker for reproducible research, with examples from the R environment},
            colorlinks=true,
            urlcolor=blue,
            linkcolor=magenta,
            pdfborder={0 0 0}}
\begin{document}

\title{An introduction to Docker for reproducible research, with examples from
the R environment}
\subtitle{}
%
%
%
%
%

\numberofauthors{1} 
%
\author{
%
%
  \alignauthor
  Carl Boettiger
  \titlenote{Corresponding author}\\
  \affaddr{\small Center for Stock Assessment Research,}\\
  \affaddr{\small 110 Shaffer Rd, Santa Cruz, CA 95050, USA}\\
  \email{\small cboettig(at)gmail.com}
}
\date{\today}

\maketitle
\begin{abstract}
As computational work becomes more and more integral to many aspects of
scientific research, computational reproducibility has become an issue
of increasing importance to computer systems researchers and domain
scientists alike. Though computational reproducibility seems more
straight forward than replicating physical experiments, the complex and
rapidly changing nature of computer environments makes being able to
reproduce and extend such work a serious challenge. In this paper, I
explore common reasons that code developed for one research project
cannot be successfully executed or extended by subsequent researchers. I
review current approaches to these issues, including virtual machines
and workflow systems, and their limitations. I then examine how the
popular emerging technology Docker combines several areas from systems
research - such as operating system virtualization, cross-platform
portability, modular re-usable elements, versioning, and a `DevOps'
philosophy, to address these challenges. I illustrate this with several
examples of Docker use with a focus on the R statistical environment.
\end{abstract}



\section{Introduction}\label{introduction}

Reproducible research has received an increasing level of attention
throughout the scientific community {[}19, 22{]} and the public at large
{[}25{]}. All steps of the scientific process, from data collection and
processing, to analyses, visualizations and conclusions depend ever more
on computation and algorithms, \emph{computational reproducibility} has
received particular attention {[}18{]}. Though in principle this
algorithmic dependence should make such research easier to reproduce --
computer codes being both more portable and potentially more precise to
exchange and run than experimental methods -- in practice this has led
to an ever larger and more complex black box that stands between what
was actually done and what is described in the literature. Crucial
scientific processes such as replicating the results, extending the
approach or testing the conclusions in other contexts, or even merely
installing the software used by the original researchers can become
immensely time-consuming if not impossible.

\subsubsection{Systems research \&
reproducibility}\label{systems-research-reproducibility}

Systems research has long concerned itself with the issues of
computational reproducibility and the technologies that can facilitate
those objectives {[}6{]}. Docker is a new but already very popular open
source tool that combines many of these approaches in a user friendly
implementation, including: (1) performing Linux container (LXC) based
operating system (OS) level virtualization, (2) portable deployment of
containers across platforms, (3) component reuse, (4) sharing, (5)
archiving, and (6) versioning of container images. While Docker's market
success has largely focused on the needs of businesses in deploying web
applications and the potential for a lightweight alternative to full
virtualization, these features have potentially important implications
for systems research in the area of scientific reproducibility. In this
paper, I seek to set these issues in the context of reproducibility
throughout the various domain sciences where computation plays an
ever-increasing role, but where researchers are largely unaware of the
concerns or technologies involved in making these computations more
reproducible, extensible, and portable to other researchers. In so
doing, I highlight elements of the Docker platform that should be of
interest to both computer systems research and domain scientists.

\subsubsection{A cultural problem}\label{a-cultural-problem}

It is worth observing from the outset that the primary barrier to
computational reproducibility in many domain sciences has nothing to do
with the technological approaches discussed here, but stems rather from
a reluctance to publish the code used in generating the results in the
first place {[}2{]}. Despite extensive evidence to the contrary
{[}14{]}, many researchers and journals continue to assume that summary
descriptions or pseudo-code provide a sufficient description of
computational methods used in data gathering, processing, simulation,
visualization, or analysis. Until such code is available in the first
place, one cannot even begin to encounter the problems that the
approaches discussed here set out to solve. As a result, few domain
researchers may be fully aware of the challenges involved in effectively
re-using published code.

A lack of requirements or incentives no doubt plays a crucial role in
discouraging sharing {[}2, 24{]}. Nevertheless, it is easy to
underestimate the significant barriers raised by a lack of familiar,
intuitive, and widely adopted tools for addressing the challenges of
computational reproducibility. Surveys and case studies find that a lack
of time, more than innate opposition to sharing, discourages researchers
from providing code {[}7, 23{]}.

\subsection{Four technical challenges}\label{four-technical-challenges}

By restricting ourselves to studies of where code has been made
available, I will sidestep for the moment the cultural challenges to
reproducibility so that I may focus on the technical ones; in
particular, those challenges for which improved tools and techniques
rather than merely norms of behavior can contribute substantially to
improved reproducibility.

Studies focusing on code that has been made available with scientific
publications regularly find the same common issues that pose substantial
barriers to reproducing the original results or building on that code
{[}4, 8, 10, 16, 27{]}, which I attempt to summarize as follows.

\subsubsection{1. \textbf{``Dependency Hell''}}\label{dependency-hell}

A recent study by researchers at the University of Arizona found that
less than 50\% of software could even be successfully built or installed
{[}4{]} and similar results are seen in an ongoing effort by other
researchers to replicate that study {[}27{]}. Installing or building
software necessary to run the code in question assumes the ability to
recreate the computational environment of the original researchers.

Differences in numerical evaluation, such as arise in floating point
arithmetic or even ambiguities in standardized programming languages
(``order-of-evaluation'' problems) can be responsible for differing
results between or even within the same computational platform {[}14{]}.
Such issues make it difficult to restrict the true dependencies of the
code to higher level environments such as that of a given scripting
language, independent of the underlying OS or even hardware itself.

\subsubsection{2. \textbf{Imprecise
documentation}}\label{imprecise-documentation}

Documentation on how to install and run code associated with published
research is another frequent barrier to replication. A study by Lapp
{[}16{]} found this impairs a researcher's ability to install and build
the software necessary, as even small holes in the documentation were
found to be major barriers, particularly for ``novices'' {[}8{]} --
where novices may be experts in nearby languages but unfamiliar with the
package managers and other tools of the language involved. This same
problem is discussed in {[}3{]}. Imprecise documentation goes well
beyond issues of the software environment itself: incomplete
documentation of parameters involved meant as few as 30\% of analyses
(\(n=34\)) using the popular software STRUCTURE could be reproduced in
the study of {[}10{]}.

\subsubsection{3. \textbf{Code rot}}\label{code-rot}

Software dependencies are not static elements, but receive regular
updates that may fix bugs, add new features or deprecate old features
(or even entire dependencies themselves). Any of these changes can
potentially change the results generated by the code. As some of these
changes may indeed resolve valid bugs or earlier problems with
underlying code, it will often be insufficient to demonstrate that
results can be reproduced when using the original versions, a problem
sometimes known as ``code rot.'' Researchers will want to know if the
results are robust to the changes. The case studies in {[}16{]} provide
examples of these problems.

\subsubsection{4. \textbf{Barriers to adoption and reuse in existing
solutions}}\label{barriers-to-adoption-and-reuse-in-existing-solutions}

Technological solutions such as workflow software, virtual machines,
continuous integration services, and best practices from software
development would address many of the issues frequently frustrating
reproducibility. However, researchers face significant barriers to entry
in learning these tools and approaches which are not part of their
typical curriculum, or lack incentives commensurate with the effort
required {[}7, 15{]}.

Though a wide variety of approaches exists to work around these
challenges, few operate on a low enough level to provide a general
solution. Clark \emph{et al.} {[}3{]} provide an excellent description
of this situation:

\begin{quote}
In scientific computing the environment was commonly managed via
Makefiles \& Unix-y hacks, or alternatively with monolithic software
like Matlab. More recently, centralized package management has provided
curated tools that work well together. But as more and more essential
functionality is built out across a variety of systems and languages,
the value -- and also the difficulty -- of coordinating multiple tools
continues to increase. Whether we are producing research results or web
services, it is becoming increasingly essential to set up new languages,
libraries, databases, and more.
\end{quote}

There are two dominant approaches to this issue of coordinating multiple
tools: Workflows and Virtual Machines (VMs).

\section{Current approaches}\label{current-approaches}

Two dominant paradigms have emerged to address these issues so far:
workflow software {[}1, 13{]} and virtual machines {[}5, 12{]}. Workflow
software provides very elegant technical solutions to the challenges of
communication between diverse software tools, capturing provenance in
graphically driven interfaces, and handling issues from versioning
dependencies to data access. Workflow solutions are often built by
well-funded collaborations between domain scientists and computer
scientists, and can be very successful in the communities within which
they receive substantial adoption. Nonetheless, most workflow systems
struggle with relatively low total adoption overall {[}5, 9{]}.

Dudley \& Butte {[}5{]} give several reasons that such comprehensive
workflow systems have not been more successful:

\begin{quote}
\begin{enumerate}
\def\labelenumi{(\roman{enumi})}
\itemsep1pt\parskip0pt\parsep0pt
\item
  efforts are not rewarded by the current academic research and funding
  environment; (ii) commercial software vendors tend to protect their
  markets through proprietary formats and interfaces; (iii)
  investigators naturally tend to want to `own' and control their
  research tools; (iv) even the most generalized software will not be
  able to meet the needs of every researcher in a field; and finally (v)
  the need to derive and publish results as quickly as possible
  precludes the often slower standards-based development path.
\end{enumerate}
\end{quote}

In short, workflow software expects a new approach to computational
research. In contrast, virtual machines (VMs) offer a more direct
approach. Since the computer Operating System (OS) already provides the
software layer responsible for coordinating all the different elements
running on the computer, the VM approach captures the OS and everything
running on it whole-cloth. To make this practical, Dudley \& Butte
{[}5{]} and Howe {[}12{]} both propose using virtual machine images that
will run on the cloud, such as Amazon's EC2 system, which is already
based upon this kind of virtualization.

Critics of the use of VMs to support reproducibility highlight that the
approach is too much of a black box and thus ill suited for
reproducibility {[}28{]}. While the approach sidesteps the need to
either install or even document the dependencies, this also makes it
more difficult for other researchers to understand, evaluate, or alter
those dependencies. Moreover, other research cannot easily build on the
virtual machine in a consistent and scalable way. If each study provided
it's own virtual machine, any pipeline combining the tools of multiple
studies would quickly become impractical or impossible to implement.

\subsection{A ``DevOps'' approach}\label{a-devops-approach}

The problems highlighted here are not unique to \emph{academic}
software, but impact software development in general. While the academic
research literature has frequently focused on the development of
workflow software dedicated to particular domains, or otherwise to the
use of virtual machines, the software development community has recently
emphasized a philosophy (rather than a particular tool), known as
\emph{Development} and Systems \emph{Operation}, or more frequently just
``DevOps.'' The approach is characterized by scripting, rather than
documenting, a description of the necessary dependencies for software to
run, usually from the Operating System (OS) on up. Clark \emph{et al.}
{[}3{]} describe the DevOps approach along with both its relevance to
reproducible research and examples of its use in the academic research
context. They identify the difficulties I have discussed so far in terms
of effective documentation:

\begin{quote}
Documentation for complex software environments is stuck between two
opposing demands. To make things easier on novice users, documentation
must explain details relevant to factors like different operating
systems. Alternatively, to save time writing and updating documentation,
developers like to abstract over such details.
\end{quote}

The authors contrast this to the DevOps approach, where dependency
documentation is \emph{scripted}:

\begin{quote}
A DevOps approach to ``documenting'' an application might consist of
providing brief descriptions of various install paths, along with
scripts or ``recipes'' that automate setup.
\end{quote}

This elegantly addresses both the demand for simplicity of use (one
executes a script instead of manually managing the environmental setup)
and comprehensiveness of implementation. Clark \emph{et al.} {[}3{]} are
careful to note that this is not so much a technological shift as a
philosophical one:

\begin{quote}
The primary shift that's required is not one of new tooling, as most
developers already have the basic tooling they need. Rather, the needed
shift is one of philosophy.
\end{quote}

Nevertheless, a growing suite of tools designed explicitly for this
purpose have rapidly replaced the use of general purpose tools (such as
Makefiles, bash scripts) to become synonymous with the DevOps
philosophy. Clark \emph{et al.} {[}3{]} reviews many of these DevOps
tools, their different roles, and their application in reproducible
research.

I focus the remainder of this paper on one of the most recent and
rapidly growing among these, called Docker, and the role it can play in
reproducible research. Docker offers several promising features for
reproducibility that go beyond the tools highlighted in {[}3{]}.
Nevertheless, my goal in focusing on this technology is not to promote a
particular solution, but to anchor the discussion of technical solutions
to reproducibility challenges in concrete examples.

\section{Docker}\label{docker}

Docker is an open source project that builds on many long-familiar
technologies from operating systems research: LXC containers,
virtualization of the OS, and a hash-based or git-like versioning and
differencing system, among others.

I introduce the most relevant concepts from Docker through the context
of the four challenges for reproducible research I have discussed above.

\subsubsection{1. Docker images: resolving `Dependency
Hell'}\label{docker-images-resolving-dependency-hell}

A Docker based approach works similarly to a virtual machine image in
addressing the dependency problem by providing other researchers with a
binary image in which all the software has already been installed,
configured and tested. (A machine image can also include all data files
necessary for the research, which may simplify the distribution of
data.)

A key difference between Docker images and other virtual machines is
that the Docker images share the Linux kernel with the host machine. For
the end user the primary consequence of this is that any Docker image
must be based on a Linux system with Linux-compatible software, which
includes (R, Python, Matlab, and most other scientific programming
needs).\footnote{Note that re-distribution of an image in which
  proprietary software has been installed will be subject to any
  relevant licensing agreement.}

Sharing the Linux kernel makes Docker much more light-weight and higher
performing than complete virtual machines -- a typical desktop computer
could run no more than a few virtual machines at once but would have no
trouble running 100's of Docker containers (a container is simply the
term for running instance of an image). This feature has made Docker
particularly attractive to industry and is largely responsible for the
immense popularity of Docker. For our purposes this is a nice bonus, but
the chief value to reproducible research lies in other aspects.

\subsubsection{2. Dockerfiles: Resolving imprecise
documentation}\label{dockerfiles-resolving-imprecise-documentation}

Though Docker images can be created interactively, this leaves little
transparent record\footnote{The situation is in fact slightly better
  than the virtual machine approach because these changes are versioned.
  Docker provides tools to inspect differences (diffs) between the
  images, and I can also roll back changes to earlier versions.} of what
software has been installed and how. Dockerfiles provide a simple script
(similar to a Makefile) that defines exactly how to build up the image,
consistent with the DevOps approach I mentioned previously.

With a syntax that is simpler than other provisioning tools (\emph{e.g.}
Chef, Puppet, Ansible) or Continuous Integration (CI) platforms
(\emph{e.g.} Travis CI, Shippable CI); users need little more than a
basic familiarity with shell scripts and a Linux distribution software
environment (\emph{e.g.} Debian-based \texttt{apt-get}) to get started
writing Dockerfiles.

This approach has many advantages:

\begin{itemize}
\item
  While machine images can be very large (many gigabytes), a Dockerfile
  is just a small plain text file that can be easily stored and shared.
\item
  Small plain text files are ideally suited for use with a version
  management system such as \texttt{subversion} or \texttt{git}, which
  can track any changes made to the \texttt{Dockerfile}
\item
  the \texttt{Dockerfile} provides a human readable summary of the
  necessary software dependencies, environmental variables and so forth
  needed to execute the code. There is little possibility of the kind of
  holes or imprecision in such a script that so frequently cause
  difficulty in manually implemented documentation of dependencies. This
  approach also avoids the burden of having to tediously document
  dependencies at the end of a project, since they are instead
  documented as they are installed by writing the \texttt{Dockerfile}.
\item
  Unlike a \texttt{Makefile} or other script, the \texttt{Dockerfile}
  includes all software dependencies down to the level of the OS, and is
  built by the Docker \texttt{build} tool, making it very unlikely that
  the resulting build will differ when being built on different
  machines. This is not to say that all builds of a Dockerfile are
  bitwise identical. In particular, builds executed later will install
  more recent versions of the same software, if available, unless the
  package managers used are explicitly configured otherwise. I address
  this issue in the next section.
\item
  It is possible to add checks and tests following the commands for
  installing the software environment, which will verify that the setup
  has been successful. This can be important in addressing the issue of
  code-rot which I discuss next.
\item
  It is straightforward for other users to extend or customize the
  resulting image by editing the script directly.
\end{itemize}

\subsubsection{3. Tackling code-rot with image
versions}\label{tackling-code-rot-with-image-versions}

As I have discussed above, changes to the dependencies, whether they are
the result of security fixes, new features, or deprecation of old
software, can break otherwise functioning code. These challenges can be
significantly reduced because Docker defines the software environment to
a particular operating system and suite of libraries, such as the Ubuntu
or Debian distribution. Such distributions use a staged release model
with \texttt{stable}, \texttt{testing} and \texttt{unstable} phases
subjected to extensive testing to catch such potential problems
{[}20{]}, while also providing regular security updates to software
within each stage. Nonetheless, this cannot completely avoid the
challenge of code-rot, particularly when it is necessary to install
software that is not (yet) available for a given distribution.

To address this concern, one must archive a binary copy of the image
used at the time the research was first performed. Docker provides a
simple utility to save an image as a portable \texttt{tarball} file that
can be read in by any other Docker installation, providing a robust way
to run the exact versions of all software involved. By testing both the
\texttt{tarball} archive and the image generated by the latest
Dockerfile, Docker provides a simple way to confirm whether or not code
rot has effected the function of a particular piece of code.

To simplify this process, Docker also supports \emph{Automated Builds}
through the Docker Hub (\href{http://hub.docker.com}{hub.docker.com}).
This acts as a kind of Continuous Integration (CI) service that verifies
the image builds correctly whenever the Dockerfile is updated,
particularly if the Dockerfile includes checks for the environment. The
Hub also provides a convenient distribution service, freely storing the
pre-built images, along with their metadata, for download and reuse by
others. The Docker Hub is a free service and an open source software
product so that users can run their own private versions of the Hub on
their own servers, for instance, if security of the data or the
longevity of the public platform is a concern.

\subsubsection{4. Barriers to adoption and
re-use}\label{barriers-to-adoption-and-re-use}

A technical solution, no matter how elegant, will be of little practical
use for reproducible research unless it is both easy to use and adapt to
the existing workflow patterns of practicing domain researchers.

Though most of the concerns I have discussed so far can be addressed
through well-designed workflow software or the use of a DevOps approach
to provisioning virtual machines by scripts, neither approach has seen
widespread adoption by domain researchers, who work primarily in a local
rather than cloud-based environment using development tools native to
their personal operating system. To gain more widespread adoption,
reproducible research technologies must make it easier, not harder, for
a researcher to perform the tasks they are already doing (before
considering any additional added benefits).

These issues are reflected both during the original research or
development phase and in any subsequent reuse. Another researcher may be
less likely to build on existing work if it can only be done by using a
particular workflow system or monolithic software platform with which
they are unfamiliar. Likewise, a user is more likely to make their own
computational environment available for reuse if it does not involve a
significant added effort in packaging and documenting {[}23{]}.

Though Docker is not immune to these challenges, it offers a interesting
example of a way forward in addressing these fundamental concerns. Here
I highlight five of these features in turn:

\begin{itemize}
\itemsep1pt\parskip0pt\parsep0pt
\item
  Local environment
\item
  Modular reuse
\item
  Portable environments
\item
  Public repository for sharing
\item
  Versioning
\end{itemize}

\subsection{Using Docker as a local development
environment}\label{using-docker-as-a-local-development-environment}

Perhaps the most important feature of a reproducible research tool is
that it be easy to learn and fit relatively seamlessly into existing
workflow patterns of domain researchers. This, more than any other
concern, can explain the relatively low uptake of previously proposed
solutions. Being a new and unfamiliar tool to most domain scientists,
Docker is far from immune to the same critique. Nevertheless, Docker
takes us several key steps towards an approach that can be easily
adopted in a research context.

While proponents of virtual machines for reproducible research propose
that these machines would be available exclusively as cloud computing
environments {[}5{]}, many researchers work \emph{locally}, that is,
primarily with software that is installed on their laptop or desktop
computer, and turning to cloud-based or other remote platforms only for
certain collaborative tasks or when the work is mature enough to need
increased computational power. Working locally allows a researcher to
rely more on the graphic interface tools for tasks such as managing
files, text editing, debugging, IDEs, or interacting with version
control systems. Scientific computing on remote machines, by contrast,
still relies largely on potentially less familiar text based command
line functions for these tasks (though web based interfaces like RStudio
Server are rapidly filling this gap).

Docker can be easily installed on most major platforms (see
\href{https://docs.docker.com/installation}{\url{https://docs.Docker.com/installation}};
On systems not already based on the Linux Kernel, such as Mac or
Windows, this is accomplished through the use of a small
VirtualBox-based VM running on the host OS called \texttt{boot2docker})
and run locally. Docker allows a user to link any directory (such as the
working directory for a particular script or project) to the running
Docker container. This allows a user to rely on the familiar tools of
the host OS for roles such as text editing, file browsing, or version
control, while still allowing code execution to occur inside the
controlled development environment of the container.

For example, one launches an interactive R console in a container that
is linked to our current working directory like so:

\begin{Shaded}
\begin{Highlighting}[]
\KeywordTok{docker} \NormalTok{run -v }\OtherTok{$(}\KeywordTok{pwd}\OtherTok{)}\NormalTok{:/ -it cboettig/rstudio /usr/bin/R}
\end{Highlighting}
\end{Shaded}

The resulting system behaves almost identically\footnote{External
  windows, such graphics windows cannot be opened directly from the
  container in this setup. Graphics would have to be saved to disk as
  raster or vector files and viewed on the host OS, until a better
  solution can be found. RStudio Server
  (\href{http://rstudio.com/products/rstudio/\#Server}{rstudio.com/products/rstudio/\#Server})
  is one way around this.} to running R on the command line of the host
OS. (Clearly a similar command could be used just as well with
interactive shells from other languages such as \texttt{ipython} or
\texttt{irb}).

\subsubsection{Using RStudio Server}\label{using-rstudio-server}

An alternative approach for working locally with familiar tools is to
leverage web-based clients such as RStudio. In the R environment, the
open source RStudio IDE provides another key component in making this
system accessible to most domain scientists. Because the RStudio IDE is
written completely with standard web languages (Javascript, CSS, and
HTML), its web-based RStudio Server looks and feels identical to its
popular desktop IDE. RStudio Server provides a way to interact with R on
a remote environment without the latency, X tunnelling, and so forth
typically involved in running R on a remote server.

RStudio server provides users a way to interact with R, the file system,
git, text editors, and graphics running on a Docker container. Users
already familiar with the popular IDE can thus benefit from the
reproducibility and portability features provided by running R in a
container environment without having to adapt to a new workflow.

To accompany this paper, I provide a Docker image for running RStudio
server, which can be launched as follows. From the terminal (or
boot2docker terminal on Mac or Windows client), run:

\begin{Shaded}
\begin{Highlighting}[]
\KeywordTok{sudo} \NormalTok{docker run -d -p 8787:8787 cboettig/ropensci}
\end{Highlighting}
\end{Shaded}

That will take a while to download the image the first time you run it.
A boot2docker user will then need to determine the ip address assigned
to boot2docker as shown below, while a Linux user can just specify
\texttt{http://localhost}.

\begin{Shaded}
\begin{Highlighting}[]
\KeywordTok{boot2docker} \NormalTok{ip}
\end{Highlighting}
\end{Shaded}

Add the port \texttt{:8787} to the end of this address and paste it into
your browser address bar, which should open to the RStudio welcome
screen. A user can now login with user/password
\texttt{rstudio/rstudio}, run R scripts, install packages, use git, and
so forth. User login and other configurations can be customized using
environmental variables; see details at
\href{https://github.com/ropensci/docker}{\url{https://github.com/ropensci/Docker}}

\subsection{Portable computation \&
sharing}\label{portable-computation-sharing}

A particular advantage of this approach is that the resulting
computational environment is immediately \emph{portable}. LXC containers
by themselves are unlikely to run in the same way, if at all, across
different machines, due to differences in networking, storage, logging
and so forth. Docker handles the packaging and execution of a container
so that it works identically across different machines, while exposing
the necessary interfaces for networking ports, volumes, and so forth.
This is useful not only for the purposes of reproducible research, where
other users may seek to reconstruct the computational environment
necessary to run the code, but is also of immediate value to the
researcher themselves. For instance, a researcher might want to execute
their code on a cloud server which has more memory or processing power
then their local machine, or would want a co-author to help debug a
particular problem. In either case, the researcher can export a snapshot
of their running container:

\begin{Shaded}
\begin{Highlighting}[]
\KeywordTok{docker} \NormalTok{export container-name }\KeywordTok{>} \NormalTok{container.tar}
\end{Highlighting}
\end{Shaded}

and then run this identical environment on the cloud or collaborators'
machine.

Sharing these images is further facilitated by the Docker Hub
technology. While Docker images tend to be much smaller than equivalent
virtual machines, moving around even 100's of gigabytes can be a
challenge. In their work on virtual machines, Dudley \emph{et al.}
{[}5{]} recommend only running these tools on cloud servers such as the
Amazon EC2 system. As most researchers still develop software locally
and may not have ready access to these resources, such a requirement
adds an additional barrier to reuse. Docker provides a convenient way to
share any image publicly or privately through the Hub after creating a
free account. Docker Hub is both a publicly available service and also
available as a separate open source platform that can be deployed on
private servers.

One can share a public copy of the image just created by using the
\texttt{Docker push} command, followed by the name of the image using
the command:

\begin{Shaded}
\begin{Highlighting}[]
\KeywordTok{docker} \NormalTok{push username/r-recommended}
\end{Highlighting}
\end{Shaded}

If a Dockerfile is made available on a public code repository such as
\href{https://github.com}{Github} or
\href{https://bitbucket.org}{Bitbucket}, the Hub can automatically build
the image whenever a change is made to the Dockerfile, making the
\texttt{push} command unnecessary. A user can update their local image
using the \texttt{Docker pull \textless{}imagename\textgreater{}}, which
downloads any changes that have since been made to the copy of the image
on the Hub.

\subsection{Re-usable modules}\label{re-usable-modules}

The approach of Linux Containers represented by Docker offers a
technical solution to what is frequently seen as the primary weakness of
the standard virtual machine approach to reproducibility - reusing and
remixing elements. To some extent this is already addressed by the
DevOps approach of Dockerfiles, providing a scripted description of the
environment that can be tweaked and altered, but also includes something
much more fundamental to Docker.

The challenge to reusing virtual machines can be summarized as ``you
can't install an image for every pipeline you want\ldots{}'' {[}28{]}.
In contrast, this is exactly how Docker containers are designed to work.
There are at least two ways in which Docker supports this kind of
extensibility.

First, Docker facilitates modular reuse by build one container on top of
another. Rather than copy a Dockerfile and then start adding more lines
to the bottom, one can declare a new Dockerfile is built on an old one
using the \texttt{FROM} directive. Here is an example Dockerfile that
adds the R statistical computing environment to a basic Ubuntu Linux
distribution.

\begin{verbatim}
FROM ubuntu:latest
RUN apt-get update 
RUN apt-get -y install r-recommended
\end{verbatim}

This acts like a software dependency; but unlike other software, a
Dockerfile must have exactly one dependency (one \texttt{FROM} line).
Note that a particular version of the dependency can be specified using
the \texttt{:} notation. One can in fact be much more precise, declaring
not only version numbers like \texttt{ubuntu:14:04}, but specific
cryptographic hashes that ensure we get the exact same image every time.

One can now build this Dockerfile and give it name, by running in the
working directory:

\begin{Shaded}
\begin{Highlighting}[]
\KeywordTok{Docker} \NormalTok{build -t username/r-recommended .}
\end{Highlighting}
\end{Shaded}

The key point here is that other researchers can easily build off this
image just created, extending our work directly, rather than having to
go back to the original image.

If this image is shared (either directly or through a Docker Hub),
another user can now build directly on our image, rather than on the
\texttt{ubuntu:latest} image used as a base in the first Dockerfile. To
do so, one simply specifies a different image in the \texttt{FROM}
directive of the Dockerfile, followed by the lines required to add any
additional software required.

\begin{verbatim}
FROM username/r-recommended 
RUN apt-get update
RUN apt-get -y install r-cran-matrix
\end{verbatim}

Though this shows only very minimal examples that add a single piece of
software in each step, clearly this approach can be particularly
powerful in building up more complex environments.

Each one acts as a building block providing just what is necessary to
run one particular service or element, and exposing just what is
necessary to link it together with other blocks. For instance, one could
have one container running a PostgreSQL database which serves data to
another container running a python environment to analyze the data:

\begin{Shaded}
\begin{Highlighting}[]
\KeywordTok{docker} \NormalTok{run -d --name db training/postgres}
\KeywordTok{docker} \NormalTok{run -d -P --link db:db training/webapp python app.py}
\end{Highlighting}
\end{Shaded}

Unlike the much more heavyweight virtual machine approach, containers
are implemented in way such that a single computer can easily run 100s
of such services each in their own container, making it easy to break
computational elements down into logically reusable chunks that come,
batteries included, with everything they need to run reproducibly. A
researcher could connect a container providing a computational
environment for a different language to this same PostgreSQL container,
and so forth.

\subsection{Versioning}\label{versioning}

In addition to version managing the Dockerfile, the images themselves
are versioned using a git-like hash system (\emph{e.g.} see
\texttt{Docker commit}, \texttt{docker push}/\texttt{Docker pull},
\texttt{docker history}, \texttt{docker diff}). Docker images and
containers have dedicated metadata specifying the date, author, parent
image, and other details (see \texttt{Docker inspect}). One can roll
back an image through the layers of history of its construction, then
build off an earlier layer, or roll back changes made interactively in a
container. For instance, here I inspect recent changes made to the
\texttt{ubuntu:latest} image:

\begin{Shaded}
\begin{Highlighting}[]
\KeywordTok{docker} \NormalTok{history ubuntu:latest}
\end{Highlighting}
\end{Shaded}

One can identify an earlier version, and roll back to that version just
by adjusting the Docker tag to match the hash of that version. For
instance:

\begin{Shaded}
\begin{Highlighting}[]
\KeywordTok{docker} \NormalTok{tag 25f ubuntu:latest}
\end{Highlighting}
\end{Shaded}

If one now inspects the history, which shows that it now begins from
this earlier point:

\begin{Shaded}
\begin{Highlighting}[]
\KeywordTok{docker} \NormalTok{history ubuntu:latest}
\end{Highlighting}
\end{Shaded}

This same feature also means that Docker can perform incremental uploads
and downloads that send only the differences between images, (just like
\texttt{git push} or \texttt{git pull} for git repositories), rather
than transfer the full image each time.

\section{Conclusions}\label{conclusions}

\subsection{Best Practices}\label{best-practices}

The effectiveness of this approach for supporting reproducible research
nonetheless depends on how each of these features are adopted and
implemented by individual researchers. I summarize a few of these
practices here:

\begin{itemize}
\item
  \emph{Use Docker containers during development}. A key feature of the
  Docker approach is the ability to mimic as closely as possible the
  current workflow and development practices of the user. Code executing
  inside a container on a local machine can appear identical to code
  running natively, but with the added benefit that one can simply
  recreate or snapshot and share the entire computational environment
  with a few simple commands. This works best if researchers set up
  their computational environment in a container from the outset of the
  project.
\item
  \emph{Write Dockerfiles instead of installing interactive sessions}.
  As we have noted already, Docker can be used in a purely interactive
  manner to record and distribute changes to a computational
  environment. However, the approach is most useful for reproducible
  research when researchers begin by defining their environment
  explicitly in the DevOps fashion by writing a Dockerfile.
\item
  \emph{Adding tests or checks to the Dockerfile}. Dockerfile commands
  need not be limited to installing software, but can also include
  execution. This can help verify that an image has build successfully
  with all the software necessary to run the research code of interest.
\item
  \emph{Use and provide appropriate base images}. Though Docker supports
  modular design, it remains up to the researchers to take advantage of
  it. An appropriate workflow might involve one Dockerfile that includes
  all the software dependencies a researcher usually uses in the course
  of their development, which can then be extended by separate Docker
  images for particular projects. Re-using existing images reduces the
  effort required to set up an environment, contributes to the
  standardization of computational environments within a field, and best
  leverages the ability of Docker's distribution system to download only
  differences.
\item
  \emph{Share Docker images and Dockerfiles}. The Docker Hub
  significantly reduces the barriers for making even large images (which
  can exceed the file size limits of journals common scientific data
  repositories such as Dryad and Figshare) readily available to other
  researchers.
\item
  \emph{Archive tarball snapshots}. Despite similar semantics to git,
  Docker's versioning system works rather differently than version
  management of code. Docker can roll back AUFS layers that have been
  added to an image, but not revert to the earlier state of a particular
  layer. In consequence, to preserve a bitwise identical snapshot of a
  container used to generate a given set of results, it is necessary to
  archive the image tarball itself -- one can not simply rely on the
  Docker history to recover an earlier state.
\end{itemize}

\subsection{Limitations and future
developments}\label{limitations-and-future-developments}

Docker has the potential to address shortcomings of certain existing
approaches to reproducible research challenges that stem from recreating
complex computational environments. Docker also provides a promising
case study in other issues. Its versioning, modular design, portable
containers, and simple interface have proven successful in industry and
could have promising implications for reproducible research in
scientific communities. Nonetheless, these advances raise questions and
challenges of their own.

\begin{itemize}
\item
  Docker does not provide complete virtualization but relies on the
  Linux kernel provided by the host. Systems research can provide
  insight on what limitations to reproducibility this introduces
  {[}11{]}.
\item
  Docker is limited to 64 bit host machines, making it impossible to run
  on older hardware (at this time).
\item
  On Mac and Windows machines Docker must still be run in a fully
  virtualized environment. Though the boot2docker tool streamlines this
  process, it remains to be seen if the performance and integration with
  the host machine's OS is sufficiently seamless or creates a barrier to
  adoption by users on of these systems.
\item
  Potential computer security issues may still need to be evaluated.
  Among other changes, future support for digitally signing Docker
  images may make it easier to build off of only trusted binaries.
\item
  Most importantly, it remains to be seen if Docker will be
  significantly adopted by any scientific research or teaching
  community.
\end{itemize}

\subsection{Further considerations}\label{further-considerations}

\subsubsection{Combining virtualization with other reproducible-research
tools}\label{combining-virtualization-with-other-reproducible-research-tools}

Using Docker containers to distribute reproducible research should be
seen as an approach that is synergistic with, rather than an alternative
to, other technical tools for ensuring computational reproducibility.
Existing tools for managing dependencies for a particular language
{[}21{]} can easily be employed within a Docker-based approach, allowing
the operating-systems level virtualization to sidestep potential issues
such as external library dependencies or conflicts with existing user
libraries. Other approaches that facilitate reproducible research also
introduce additional software dependencies and possible points of
failure {[}7{]}. One example includes dynamic documents {[}17, 22, 26{]}
which embed the code required to re-generate the results within the
manuscript. As a result, it is necessary to package the appropriate
typesetting libaries (\emph{e.g.} \LaTeX) along with the code libaries
such that the document executes successfully for different researchers
and platforms.

\subsubsection{Impacting cultural
norms?}\label{impacting-cultural-norms}

I noted at the outset that cultural expectations responsible for a lack
of code sharing practices in many fields are a far more extensive
primary barrier to reproducibility than the technical barriers discussed
here. Nevertheless, it may be worth considering how solutions to these
technical barriers can influence the cultural landscape as well. Many
researchers may be reluctant to publish code today because they fear a
it will be primarily a one-way street: more technical savvy researchers
then themselves can benefit from their hard work, while they may not
benefit from the work produced by others. Lowering the technical
barriers to reuse provides immediate practical benefits that make this
exchange into a more balanced, two-way street. Another concern is that
the difficulty imposed in preparing code to be shared, such as providing
even semi-adequate documentation or support for other users to be able
to install and run it in the first place is too high {[}23{]}. Thus,
lowering these barriers to re-use through the appropriate infrastructure
may also reduce certain cultural barriers to sharing.

\section{Acknowledgements}\label{acknowledgements}

CB acknowledges support from NSF grant DBI-1306697, and also from the
Sloan Foundation support through the rOpenSci project. CB also wishes to
thank Scott Chamberlain, Dirk Eddelbuettel, Rich FitzJohn, Yihui Xie,
Titus Brown, John Stanton-Geddes and many others for helpful discussions
about reproducibility, virtualization, and Docker that have helped shape
this manuscript.

\section{References}\label{references}

{[}1{]}Altintas, I. et al. 2004. Kepler: an extensible system for design
and execution of scientific workflows. \emph{Proceedings. 16th
international conference on scientific and statistical database
management, 2004.} (2004).

{[}2{]}Barnes, N. 2010. Publish your computer code: it is good enough.
\emph{Nature}. 467, 7317 (Oct. 2010), 753--753.

{[}3{]}Clark, D. et al. 2014. BCE: Berkeley's Common Scientific Compute
Environment for Research and Education. \emph{Proceedings of the 13th
Python in Science Conference (SciPy 2014)}. (2014).

{[}4{]}Collberg, C. et al. 2014. \emph{Measuring Reproducibility in
Computer Systems Research}.

{[}5{]}Dudley, J.T. and Butte, A.J. 2010. In silico research in the era
of cloud computing. \emph{Nat Biotechnol}. 28, 11 (Nov. 2010),
1181--1185.

{[}6{]}Eide, E. 2010. Toward Replayable Research in Networking and
Systems. \emph{Archive '10, the nSF workshop on archiving experiments to
raise scientific standards} (2010).

{[}7{]}FitzJohn, R. et al. 2014. Reproducible research is still a
challenge. \url{http://ropensci.org/blog/2014/06/09/reproducibility/}.

{[}8{]}Garijo, D. et al. 2013. Quantifying reproducibility in
computational biology: The case of the tuberculosis drugome.
\emph{\(\lbrace\)PLoS\(\rbrace\) \(\lbrace\)ONE\(\rbrace\)}. 8, 11 (Nov.
2013), e80278.

{[}9{]}Gil, Y. et al. 2007. Examining the challenges of scientific
workflows. \emph{Computer}. 40, 12 (2007), 24--32.

{[}10{]}Gilbert, K.J. et al. 2012. Recommendations for utilizing and
reporting population genetic analyses: the reproducibility of genetic
clustering using the program structure. \emph{Mol Ecol}. 21, 20 (Sep.
2012), 4925--4930.

{[}11{]}Harji, A.S. et al. 2013. Our Troubles with Linux Kernel Upgrades
and Why You Should Care. \emph{ACM SIGOPS Operating Systems Review}. 47,
2 (2013), 66--72.

{[}12{]}Howe, B. 2012. Virtual appliances, cloud computing, and
reproducible research. \emph{Computing in Science \& Engineering}. 14, 4
(Jul. 2012), 36--41.

{[}13{]}Hull, D. et al. 2006. Taverna: a tool for building and running
workflows of services. \emph{Nucleic Acids Research}. 34, Web Server
(Jul. 2006), W729--W732.

{[}14{]}Ince, D.C. et al. 2012. The case for open computer programs.
\emph{Nature}. 482, 7386 (Feb. 2012), 485--488.

{[}15{]}Joppa, L.N. et al. 2013. Troubling Trends in Scientific Software
Use. \emph{Science (New York, N.Y.)}. 340, 6134 (May 2013), 814--815.

{[}16{]}Lapp, Hilmar 2014. Reproducibility / repeatability bigThink
(with tweets) @hlapp. \emph{Storify}.
\url{http://storify.com/hlapp/reproducibility-repeatability-bigthink}.

{[}17{]}Leisch, F. 2002. Sweave: Dynamic Generation of Statistical
Reports Using Literate Data Analysis. \emph{Compstat}. W. Härdle and B.
Rönz, eds. Physica-Verlag HD.

{[}18{]}Merali, Z. 2010. Computational science: ...Error. \emph{Nature}.
467, 7317 (Oct. 2010), 775--777.

{[}19{]}Nature Editors 2012. Must try harder. \emph{Nature}. 483, 7391
(Mar. 2012), 509--509.

{[}20{]}Ooms, J. 2013. Possible directions for improving dependency
versioning in r. \emph{arXiv.org}.
\url{http://arxiv.org/abs/1303.2140v2}.

{[}21{]}Ooms, J. 2014. The openCPU system: Towards a universal interface
for scientific computing through separation of concerns.
\emph{arXiv.org}. \url{http://arxiv.org/abs/1406.4806}.

{[}22{]}Peng, R.D. 2011. Reproducible research in computational science.
\emph{Science}. 334, 6060 (Dec. 2011), 1226--1227.

{[}23{]}Stodden, V. 2010. The scientific method in practice:
Reproducibility in the computational sciences. \emph{SSRN Journal}.
(2010).

{[}24{]}Stodden, V. et al. 2013. Setting the Default to Reproducible.
(2013), 1--19.

{[}25{]}The Economist 2013. How science goes wrong. \emph{The
Economist}.
\url{http://www.economist.com/news/leaders/21588069-scientific-research-has-changed-world-now-it-needs-change-itself-how-science-goes-wrong}.

{[}26{]}Xie, Y. 2013. \emph{Dynamic documents with R and knitr}.
Chapman; Hall/CRC.

{[}27{]}2014. Examining reproducibility in computer science.
\url{http://cs.brown.edu/~sk/Memos/Examining-Reproducibility/}.

{[}28{]}2012. Mick Watson on Twitter: @ewanbirney @pathogenomenick
@ctitusbrown you can't install an image for every pipeline you want...
\url{https://twitter.com/BioMickWatson/status/265037994526928896}.

\end{document}